# Metalorganic chemical vapor deposition of hexagonal boron nitride on (001) sapphire substrates for thermal neutron detector applications

K. Ahmed,[1,a)] R. Dahal,[1,] A. Weltz,[2] James J.-Q. Lu,[1] Y. Danon,[2] and I. B. Bhat[1]

[1]*Department of Electrical, Computer, and Systems Engineering, Rensselaer Polytechnic Institute, Troy, NY 12180, USA*
[2]*Department of Mechanical, Aerospace and Nuclear Engineering, Rensselaer Polytechnic Institute, Troy, NY 12180, USA*

This paper reports on the growth and characterization of hexagonal boron nitride (hBN) and its use for solid-state thermal neutron detection. The hBN epilayers were grown by metalorganic chemical vapor deposition on sapphire substrates at a temperature of 1350 °C. X-ray diffraction peak from the (002) hBN plane at a 2θ angle of 26.7° exhibited the c-lattice constant of 6.66 Å for these films. A strong peak corresponding to the high frequency Raman active mode of hBN was found for the films at 1370.5 cm$^{-1}$. hBN-based solid-state neutron detectors were fabricated and tested with a metal-semiconductor-metal configuration with an electrode spacing of 1 mm and hBN thickness of 2.5 µm. Fabricated detectors showed strong response to deep UV light as well. An intrinsic thermal neutron detection efficiency of 0.86% was measured, which is close to the theoretically expected efficiency of 0.87%. These results demonstrate that epitaxial hBN films are promising for thermal neutron detection applications.

$^3$He proportional counters are considered the gold standard for thermal neutron detection due to their high intrinsic thermal neutron detection efficiency (η) and exceptionally low gamma sensitivity [1]. However, the material costs associated with $^3$He production are exceedingly high [2]. Additionally, $^3$He gas filled detectors require a rather large bias voltage and are bulky, which limit their portability [3-8]. Solid-state neutron detectors (SSNDs) can overcome the limitations of $^3$He neutron detectors due to their compact nature, reduced bias voltage requirement, fast timing characteristics, chip level integration with devices like cell phones, and significantly reduced material costs [8, 9]. The development of SSNDs with hBN is promising due to the large thermal neutron capture cross section of $^{10}$B, a constituent of hBN. $^{10}$B has a thermal neutron capture cross section of ~3840 barns at 0.0253 eV and makes up 19.9% of natural boron [5]. Additionally, hBN neutron detectors are referred to as homogenous SSNDs since both the neutron conversion and charge collection happen in the same material (i.e., hBN). A homogeneous SSND has the potential to provide higher η and an improved pulse height spectrum compared to heterogeneous counterparts, which employ a separate neutron conversion region (typically $^{10}$B or $^6$Li) and charge-collecting region (e.g., Si). Furthermore, hBN-based SSNDs are theoretically more robust than Si-based SSNDs because hBN is less susceptible to damage from energetic charge particles due to its wide bandgap. Additionally, the low atomic numbers of B (5) and N (7) ensure that hBN detectors are relatively insensitive to gamma rays, which is an important characteristic of a good neutron detector [4, 5]. When an incident thermal neutron is absorbed by a $^{10}$B atom in an hBN SSND, the following reactions can occur [4, 5]:

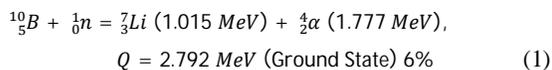

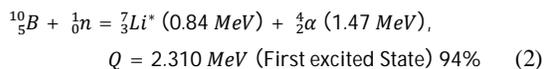

The resulting energetic, charged particles, α particles and $^7$Li ions, have a small range of ~5 µm and ~2 µm in hBN, respectively [4, 5]. These particles lose energy in hBN and produce electron-hole pairs (EHPs). These EHPs are collected under an applied electric field in the hBN SSND, which produce a detection signal referring to the presence of the thermal neutron.

This work focuses on the development of epitaxial growth of hBN on sapphire substrates using a cold wall metalorganic chemical vapor deposition (MOCVD) system, and the characterization of the crystalline and bonding properties of such hBN films. Furthermore, the hBN films were used to fabricate metal-semiconductor-metal (MSM) devices. These devices have charge transport along the a-axis of hBN, which is the highest charge carrier mobility path in hBN. In order to characterize these MSM devices, their responses were measured under deep UV (DUV) excitation. Additionally, alpha and neutron responses were collected using a 1.0 kBq $^{241}$Am source and a moderated $^{252}$Cf spontaneous fission source, respectively. A η of 0.86% was estimated for a 2.5 µm thick hBN based MSM structure SSND, indicating a high thermal neutron conversion efficiency of ~98%. Such efficiency measurements were not reported for hBN SSNDs (with charge collection along the a-axis of hBN) developed elsewhere, which had much thinner hBN layer and thus much lower theoretical maximum efficiency [4, 5]. Scaling the detection area up of a SSND is possible if both the leakage current ($I_{dev}$) and device capacitance ($C_{dev}$) are small [8]. hBN MSM detectors have orders of magnitude smaller $I_{dev}$ (because of high resistivity of hBN) and orders of magnitude smaller $C_{dev}$, as compared to Si-based SSNDs. Hence, these hBN detectors could be scaled to a much larger detection area than that of Si-based SSNDs.

A custom-built cold wall MOCVD system was used for the epitaxial growth of hBN films at growth temperatures above 1300 °C. Silicon carbide coated graphite susceptors were used for the high temperature growths, and the susceptors were heated with an RF coil utilizing the induction heating system. hBN growth was performed on sapphire substrates using triethylboron (TEB) and ammonia ($NH_3$) as the precursors for B and N, respectively. A natural TEB source (19.9% $^{10}$B and 80.1% $^{11}$B) was used. All samples were grown with a reactor pressure of 100 torr. Hydrogen ($H_2$) was used as the carrier gas. A large lattice mismatch exists between sapphire and hBN (in-plane lattice constants are 4.758 Å and 2.5 Å for sapphire and hBN, respectively) [10]. Previous studies of MOCVD growth of $sp^2$-BN phases (i.e., hBN and rhombohedral BN, rBN) employed a sapphire nitridation step (at the $sp^2$-BN growth temperature) to form a thin AlN (a-lattice constant of 3.11 Å) buffer layer to reduce the lattice mismatch [11, 12]. In this work, a thin and amorphous nitridated layer was formed as the buffer layer by annealing the sapphire substrate at 850 °C with a $NH_3$ flux of 55 µmol/s in $H_2$ environment for 10 minutes. The temperature ramp-up from the nitridation temperature (850 °C) to the hBN epitaxial growth temperature (1350 °C) was performed in 5 minutes in the $H_2$ atmosphere with the flow of both precursors turned off. The hBN epilayers were then grown at 1350 °C with a V/III ratio of 300. The precursor fluxes were 0.5 µmol/s for TEB and 150 µmol/s for $NH_3$. A fixed $H_2$ flow of 2 slm was employed for the entire growth duration.

In order to determine the crystalline quality of the deposited hBN films, X-ray diffraction (XRD) measurements were performed. A Bruker D8-Discover X-Ray Diffractometer (with Cu Kα radiation of 1.541 Å wavelength) was used for the XRD measurements. An XRD θ-2θ scan of a representative hBN film of 0.3 µm thickness is shown in Fig. 1 (a). The θ-2θ scan revealed a peak corresponding to the (002) plane of hBN at a 2θ angle of 26.7° with a full width at half maximum (FWHM) value of 0.25°. The other peak at 41.7° corresponds to the (006) plane of the sapphire substrate. Absence of any peak around 36° (corresponding to (002) AlN plane) denotes that the nitridated buffer layer is thin and amorphous. The hBN peak at 26.7° corresponds to the c-lattice constant of 6.66 Å, which matches the c-lattice constant (6.66 Å) of bulk hBN [13-15]. An XRD rocking curve (ω scan) of the (002) diffraction peak of the same hBN film has a FWHM of 0.38°, as shown in Fig. 1 (b). The observed linewidth is broader than some reported hBN films, which have a FWHM of ~385 arcsec [3, 12]. As with other III-Nitrides (e.g., GaN), the FWHM of the XRD rocking curve corresponding to the hBN films depends on the dislocation density along with stacking faults, and misalignment between hBN layers [5]. Thick hBN films (with thickness above 1 µm) were grown using a much lower V/III ratio of 125 and a much higher growth rate (a factor of 5 higher than that of the 0.3 µm film). With increasing thickness for these thick films, (002) hBN peaks become broader and shift to lower 2θ angles compared to 26.7°, indicating poorer crystalline quality and the formation of turbostratic hBN, a disordered phase of hBN. hBN films with the thickness of 2.5 µm and 15 µm exhibit broad (002) hBN peaks at 26.2° and 26°, respectively (not shown here).

Raman spectroscopy provides complementary evidence of the hBN phase of the grown films. Figure 1 (c) shows a Raman spectrum of the same 0.3 µm hBN film. Each Raman spectrum was collected using a Witec Alpha 300R confocal Raman imaging system with a 532 nm laser excitation source at room temperature. hBN films showed a Raman peak at 1370.5 $cm^{-1}$ with a FWHM of 25 $cm^{-1}$. This peak corresponds to the $E_{2g}$ symmetry vibrations in hBN due to in-plane atomic displacements. The Raman peak is almost at the same position but with a higher FWHM when compared to bulk hBN crystals grown using the flux method with a nickel and chromium solvent mixture (peak at 1370 $cm^{-1}$ with a FWHM of 9 $cm^{-1}$) [17]. Similar Raman peaks were observed for hBN nanoplates synthesized by a combination of combustion process and subsequent annealing under nitrogen atmospheric conditions (peak at 1369 $cm^{-1}$ with a FWHM of 20 $cm^{-1}$) [18].

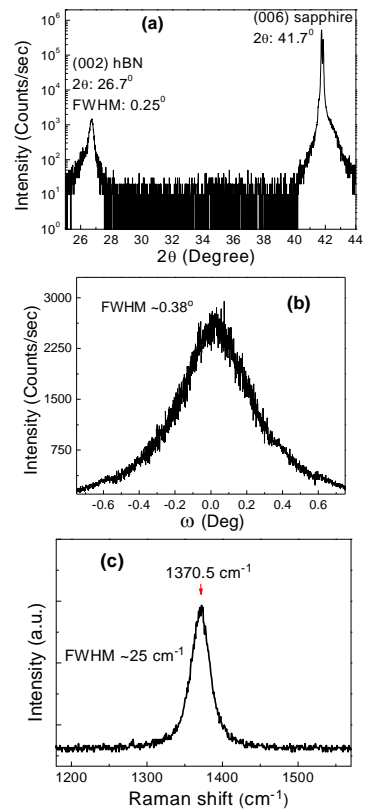

Fig. 1. (a) θ-2θ XRD scan, (b) XRD rocking curve of (002) reflection, and (c) Raman spectrum of a 0.3 µm hBN film.

hBN MSM devices were fabricated using a complete dry processing. An optical image and a corresponding schematic of an MSM device fabricated using a 2.5 µm hBN epilayer are shown in Fig. 2 (a) and (b), respectively. The effective device area is 0.4 $cm^2$, and the electrode spacing of the MSM device is 1 mm. Etching of the hBN epilayers was done by inductively coupled plasma-reactive ion etching (ICP- RIE) using $SF_6$ plasma. A Trion Phantom III Reactive Ion Etch System was used for the etching procedure. E-beam evaporation was used to deposit metal (Ni) to form metal contacts.

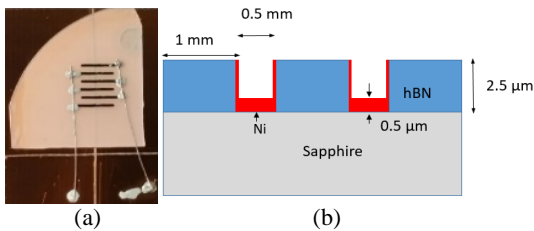

Fig. 2. (a) Optical image and (b) schematic of an MSM device with 2.5 µm thick hBN. Device area is 0.4 cm$^2$.

I-V characteristics of the fabricated MSM device (with 2.5 µm thick hBN) are shown in Fig. 3 (a), which incorporates both the dark current and current under DUV excitation. A UVP pen ray lamp was used, which has an intensity of 160 µW/cm$^2$ at a distance of 0.75 inches (corresponding to the DUV wavelength of 184.9 nm). This MSM device structure is symmetrical and its photocurrent is equal in magnitude for equal bias voltages of opposite polarities. Figure 3 (b) shows the photocurrent decay kinetics for the same device at a bias voltage of 150 V, which confirm that there are no significant persistent photoconductivity (PPC) effects in the device.

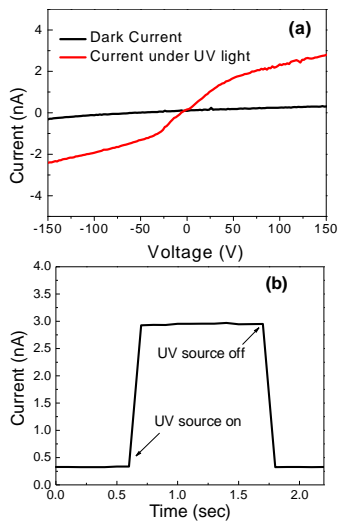

Fig. 3. (a) I-V characteristics and (b) photocurrent decay kinetics (with 150 V applied voltage) of an MSM device. This 0.4 cm$^2$ area device has 2.5 µm thick hBN.

The performance of the same MSM device (with 2.5 µm thick hBN) was tested using a 1.0 kBq $^{241}$Am alpha source ($E_\alpha$= 5.5 MeV). A bias voltage of 750 V was applied to the MSM device. Two pulse height spectra were obtained for the device with the alpha source placed directly on the device and subsequently with the alpha source removed (to measure background noise). Figure 4 (a) shows these pulse height spectra, which demonstrates that the hBN MSM device is sensitive to alpha particles.

The neutron response of the same MSM device was measured by exposing the device to a $^{252}$Cf source placed in a high density polyethylene (HDPE) moderator housing. The moderator reduces the energy of the fast neutrons emitted by the $^{252}$Cf source to that of thermal neutrons. The applied bias voltage was 750 V. One pulse height spectrum was obtained with the MSM device placed close to the front of moderator housing. Another pulse height spectrum measurement was done with the $^{252}$Cf source removed from the moderator housing to obtain the electronic noise level. The neutron and background pulse height spectra are shown in Fig. 4 (b), which demonstrate that the hBN MSM device detects neutrons.

These responses were reproducible, and the η of the device was measured by exposing the device to the same moderated neutron source. Figure 4 (c) illustrates three different pulse height spectra measured for the MSM device biased with 750 V, and the device was placed 8 cm away from the front of the moderator housing for all the measurements. The thermal neutron flux is well-calibrated at this position and was previously characterized using gold foil activation method. The thermal neutron flux at this position was calculated to be 380 n/cm$^2$-s at the time of the measurement. One measurement was performed with the $^{252}$Cf source placed in the moderator housing. The second measurement was done with the $^{252}$Cf source placed in the moderator housing and a 2 mm thick cadmium (Cd) shield placed between the detector and the moderator. The moderated neutron source emits fast neutrons and gamma rays along with the thermal neutrons. Cd was used to block thermal neutrons and allow gamma rays and fast neutrons to pass. Hence, the counts recorded in the second measurement was subtracted from the counts recorded in the first measurement to determine the counts registered by the thermal neutrons. The third measurement was performed with the neutron source removed from the moderator housing to determine the background noise. The detection efficiency of the 0.4 cm$^2$ area MSM device was estimated to be 0.86%. All the pulse height spectra related to alpha and neutron response were collected using a preamplifier (ORTEC 142PC), a pulse shaping amplifier (ORTEC 672), and a multichannel analyzer (ORTEC ASPEC-927). The pulse shaping amplifier had a gain of 300 and a shaping time of 3 µs for the alpha response measurement shown in Fig. 4 (a), and a gain of 150 and a shaping time of 3 µs for the neutron response measurements shown in Fig. 4 (b) and (c). Different gain values of the shaping amplifier and different measurement durations caused different background noise levels of Fig. 4 (a), (b), and (c).

When a thermal neutron interacts with a sufficiently thick hBN layer in an hBN SSND, the resulting charged particles deposit all of their energy in hBN. Ideally, every neutron interaction will result in full charge deposition and collection, so each pulse height will correspond to the Q value of the reaction (2.31 MeV). However, charge trapping and edge effects result in less than ideal charge collection. Neutron interaction probability (p) of $^{10}$B can be calculated from the relation, $p = 1 - e^{-\Sigma t} = 1 - e^{-\sigma N t}$, where Σ is the macroscopic cross section of $^{10}$B, σ is the microscopic cross section of $^{10}$B, N is the atomic number density of $^{10}$B, and t is the thickness of $^{10}$B. Assumed values are Σ=35.01 cm$^{-1}$ and σ=3450 × 10$^{-24}$ cm$^2$ for the moderated neutrons in this work. A layer of natural hBN (with 19.9% $^{10}$B) with a thickness of 2.5 µm should have a neutron interaction probability of 0.87%. Hence, the maximum intrinsic thermal neutron detection efficiency ($\eta_{max}$) for a 2.5 µm thick natural hBN device is 0.87%. The measured η of the MSM device with 2.5 µm thick natural hBN is 0.86%, which indicates an excellent neutron conversion efficiency (~98%) with an applied electric field of 7500 V/cm. A 2.5 µm thick hBN based MSM device would possess a η of 4.3% if enriched TEB source (with ~100% $^{10}$B) were used instead of the natural TEB source for hBN growth. MSM devices with thicker hBN would give higher η. η was not estimated for previously reported hBN MSM SSNDs which incorporated much thinner (0.3 µm) natural hBN and had a theoretical $\eta_{max}$ of ~0.1% [4, 5]. Wet processing (i.e., photolithography) was used for the fabrication of those SSNDs, whereas a complete dry processing was employed for the fabrication of the 2.5 µm hBN based SSND in this work. hBN SSND reported here has very low $I_{dev}$ and $C_{dev}$, and scalable large detection area is achievable with this detector.

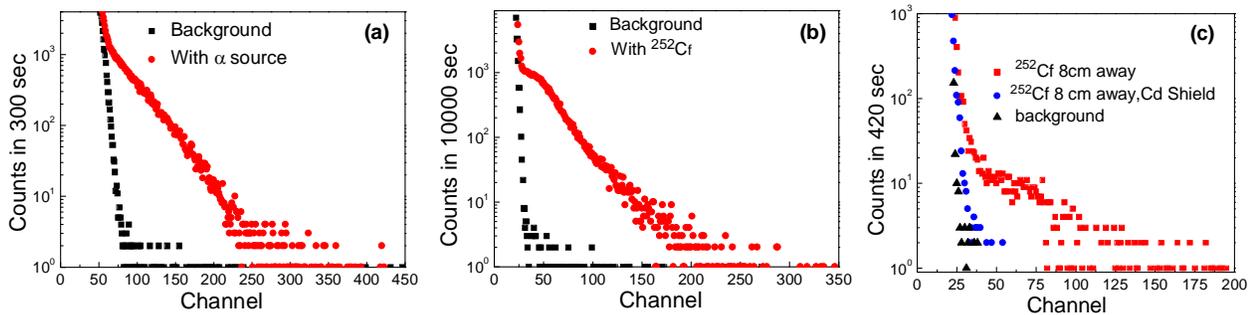

Fig. 4. (a) Alpha and background response and (b) neutron and background response of an MSM device. The MSM device has 2.5 µm thick hBN film, an area of 0.4 cm$^2$, and an electrode spacing of 1 mm. (c) Pulse height distribution of the same device for intrinsic thermal neutron detection efficiency measurement. Applied bias was 750V for all the measurements.

Good quality hBN epilayers were grown on sapphire substrates by MOCVD in order to produce homogeneous MSM solid-state thermal neutron detectors. The hBN phase of the grown films was confirmed by XRD and Raman spectroscopy measurements. The hBN MSM devices show good response to DUV light, alpha particles, and thermal neutrons. An MSM device with an hBN thickness of 2.5 µm and a device area of 0.4 cm$^2$ was exposed to a moderated $^{252}$Cf source, and the corresponding pulse height spectra show an intrinsic thermal neutron detection efficiency of 0.86%, which closely matches the theoretical maximum efficiency for such thick film, indicating an excellent effective neutron conversion efficiency of ~98%. Achieving high neutron conversion efficiency and a measureable intrinsic thermal neutron efficiency of an hBN SSND is a significant step toward the development of a high-efficiency, large area, and low-cost homogeneous SSND.

**ACKNOWLEDGMENTS**

The authors would like to thank the staff of the Rensselaer Polytechnic Institute Micro and Nano Fabrication Clean Room for their support and input. This research work is supported by DHS DNDO under competitively awarded grants ECCS-1348269 and 2013-DN-077-ER001.